\documentclass[structabstract]{aa}
\usepackage{graphics,graphicx}
\usepackage{amssymb}
\usepackage{rotating}
\usepackage{natbib}
\usepackage{savesym}
\usepackage{amsmath}
\savesymbol{iint}
\usepackage{txfonts}
\restoresymbol{TXF}{iint}

\bibpunct{(}{)}{;}{a}{}{,}

\begin{document}
\def\braces#1{[#1]}

\title{Multiwavelength study of the newly confirmed supernova remnant MCSNR~J0527$-$7104 in the Large Magellanic Cloud}
 % \thanks{}

 % \subtitle{}

\author{P. J. Kavanagh\inst{1} \and M. Sasaki\inst{1} \and S. D. Points\inst{2} \and M. D. Filipovi\'{c}\inst{3} \and P. Maggi\inst{4} \and \\ L. M. Bozzetto\inst{3} \and E. J. Crawford\inst{3} \and F. Haberl\inst{4} \and W. Pietsch\inst{4}}

\offprints{P. J. Kavanagh, \email{kavanagh@astro.uni-tuebingen.de}}

\institute{Institut f\"{u}r Astronomie und Astrophysik, Kepler Center for Astro and Particle Physics, Eberhard Karls Universit\"{a}t, 72076 T\"{u}bingen, Germany 
    \and Cerro Tololo Inter-American Observatory, Casilla 603, La Serena, Chile
    \and University of Western Sydney, Locked Bag 1797, Penrith NSW 2751, Australia
  \and Max-Planck-Institut f\"{u}r extraterrestrische Physik, Giessenbachstra\ss{}e, 85748 Garching, Germany}

\date{Received - / Accepted -}

\abstract{The Large Magellanic Cloud (LMC) hosts a rich and varied population of supernova remnants (SNRs). Optical, X-ray, and radio observations are required to identify these SNRs, as well as to ascertain the various processes responsible for the large array of physical characteristics observed.}{In this paper we attempted to confirm the candidate SNR [HP99]~1234, identified in X-rays with \textit{ROSAT}, as a true SNR by supplementing these X-ray data with optical and radio observations.}{Optical data from the Magellanic Cloud Emission Line Survey (MCELS) and new radio data from the Molonglo Observatory Synthesis Telescope (MOST), in addition to the \textit{ROSAT} X-ray data, were used to perform a multiwavelength morphological analysis of this candidate SNR.}{An approximately ellipsoidal shell of enhanced [\ion{S}{ii}] emission, typical of an SNR ([\ion{S}{ii}]/H$\alpha >$ 0.4), was detected in the optical. This enhancement is positionally coincident with faint radio emission at $\lambda$ = 36 cm. Using the available data we estimated the size of the remnant to be $\sim5.1\arcmin \times\ 4.0\arcmin$ ($\sim75$ pc $\times 59$ pc). However, the measurement along the major-axis was somewhat uncertain due to a lack of optical and radio emission at its extremities and the poor resolution of the X-ray data. Assuming this SNR is in the Sedov phase and adopting the ambient mass density of $1.2 \times 10^{-25}$ g cm$^{-3}$ measured in a nearby \ion{H}{ii} region, an age estimate of $\sim$25~kyr was calculated for a canonical initial explosion energy of $10^{51}$ erg. However, this age estimate should be treated cautiously due to uncertainties on the adopted parameters. Analysis of the local stellar population suggested a type Ia event as a precursor to this SNR, however, a core-collapse mechanism could not be ruled out due to the possibility of the progenitor being a runaway massive star.}{With the detection of X-ray, radio and significant optical line emission with enhanced [\ion{S}{ii}], this object was confirmed as an SNR to which we assign the identifier MCSNR~J0527$-$7104.}

\titlerunning{Multiwavelength study of MCSNR~J0527$-$7104 in the LMC}
\maketitle 

\section{Introduction}
Supernova remnants (SNRs) are of vital importance to the physical and chemical evolution of the interstellar medium (ISM). The expanding shells of the remnants impart kinetic energy to the surrounding ISM, as well as enriching it with the metals synthesized in the cores of their progenitor stars. Thus, an understanding of these objects is crucial to the understanding of star formation and matter recycling in galaxies. Unfortunately, due to SNRs being located primarily in the Galactic disk, studies of these objects in the Galaxy are hindered by high foreground absorption. Hence, we must look to nearby galaxies to perform unbiased studies of SNRs and SNR populations.\\

\par The \object{Large Magellanic Cloud} (LMC) is, for a variety of reasons, the best galaxy in the Local Group for the study of SNRs. At a distance of 50 kpc \citep{diBen2008} it is sufficiently close that its stellar population and diffuse structure is resolved in most wavelength regimes. In addition, the \object{LMC} is almost face-on \citep[inclination angle of $\sim 30^{\circ}$-$40^{\circ}$,][]{Marel2001,Nikolaev2004} making the entire population of SNRs available for study. The modest extinction in the line of sight (average Galactic foreground $N_{\rm{H}} \approx 7 \times 10^{20}\ \rm{cm}^{-2}$) means optical and X-ray observations of SNRs are only slightly affected by foreground absorption, whereas its location in one of the coldest parts of the radio sky \citep{Haynes1991} allows for improved radio observations without interference from Galactic emission. There are over 50 confirmed SNRs in the LMC \citep{Badenes2010,Klimek2010} with a further $\sim20$ SNR candidates \citep{Payne2008}.  However, continuing studies with new and archival optical, radio, and X-ray observations strive to add to the number of these SNRs and candidate SNRs. In addition, ever improving multiwavelength observational data allows the characterisation of the physical processes responsible for this rich and varied population. Recent works in this area include \citet{Maggi2012,Grondin2012}.
%\citet{Maggi2012,Haberl2012,deHorta2012,Grondin2012,Bozzetto2012}.

\begin{figure*}[!ht]
\begin{center}
\resizebox{\hsize}{!}{\includegraphics{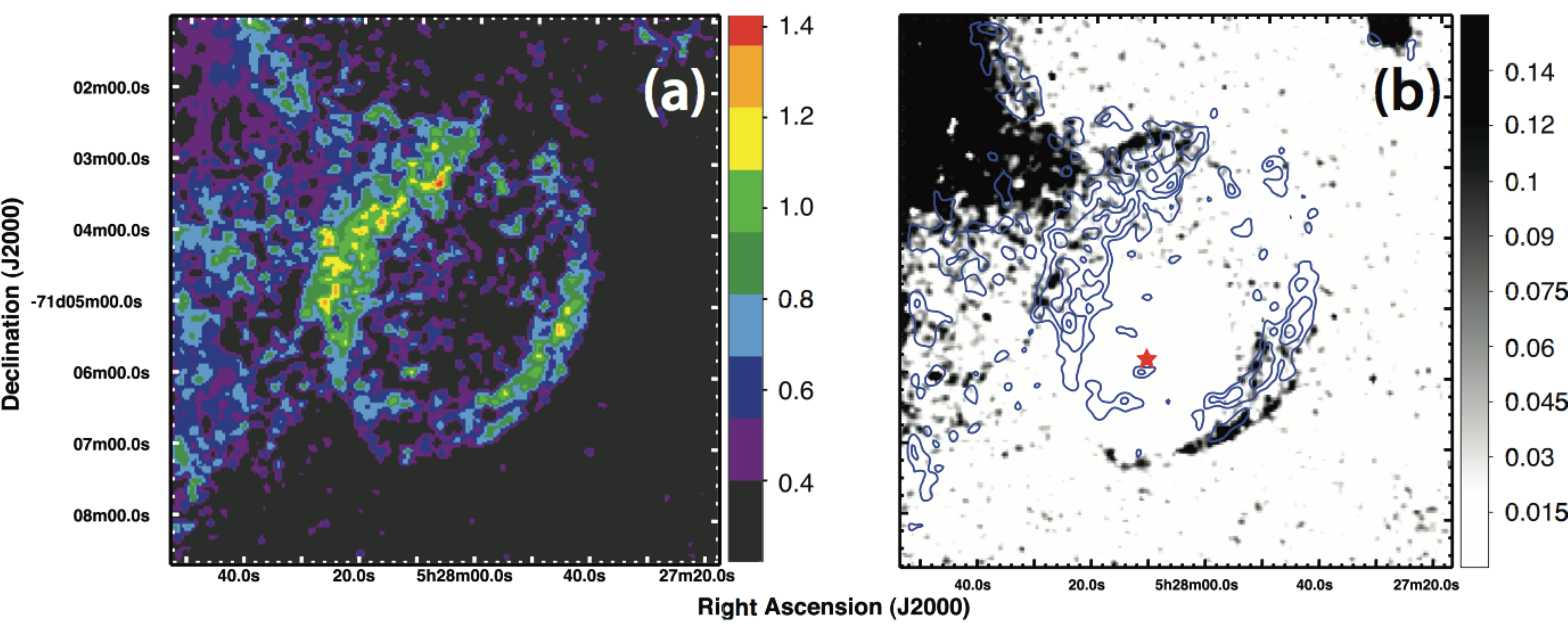}}
\caption{MCELS derived images of MCSNR~J0527$-$7104. (a) [\ion{S}{ii}]/H$\alpha$ ratio map with the colour scale adjusted so only regions of enhanced [\ion{S}{ii}] are visible. The SNR is clearly seen with two prominent arcs of enhanced [\ion{S}{ii}] located NE and SW of the centre. The emission to the east of the SNR is due to the N~206 \ion{H}{ii} complex. (b) Continuum subtracted [\ion{O}{iii}] image of MCSNR~J0527$-$7104.  [\ion{S}{ii}]/H$\alpha$ contours are overlaid, emphasising regions where the ratio is clearly enhanced, ranging from 0.67 to the maximal value with intermediate levels of 25, 50, and 75 \% of the maximum. The grayscale is in units of $10^{-15}$ erg cm$^{-2}$ s$^{-1}$. The location of the bright foreground source HD 36877, which was masked for the continuum subtraction, is indicated by the red star. Each image has been smoothed using a $3\times3$ pixel Gaussian filter.}
\label{mcels}
\end{center}
\end{figure*}

\par Various SNR classification criteria exist in the literature. In this paper we adopt the criteria of the Magellanic Cloud Supernova Remnant (MCSNR) Database\footnote{See \url{http://www.mcsnr.org/about.aspx}}. In this classification system, an SNR must satisfy at least two of the following three observational criteria: significant H$\alpha$, [\ion{S}{ii}], and/or [\ion{O}{iii}] line emission with an [\ion{S}{ii}]/H$\alpha$ flux ratio $>0.4$ \citep{Matt1973,Fesen1985}; extended non-thermal radio emission; and extended thermal X-ray emission. The \textit{ROSAT} PSPC catalogue of X-ray sources in the LMC \citep[][hereafter HP99]{Haberl1999} contains \object{[HP99]~1234} at a J2000 position of RA = 05h27m57.2s and DEC = $-$71d04m30s. It was classified as a candidate SNR in HP99 based on its X-ray hardness ratios and extent. However, without optical or radio detections to supplement the X-ray identification, this source has remained mired in the `candidate SNR' category. \\

\par In this paper we report the identification of enhanced [\ion{S}{ii}] emission typical of an SNR in the vicinity of [HP99]~1234 using data from the Magellanic Cloud Emission Line Survey (MCELS). In addition, we detected faint radio emission positionally coincident with the [\ion{S}{ii}] enhancement using observations by the Molonglo Synthesis Telescope (MOST) at $\lambda = 36$ cm. Hence, [HP99]~1234 appears to satisfy all three criteria for SNR classification and is, thus, confirmed as such. We will hereafter refer to this source as \object{MCSNR~J0527$-$7104} (adopting the MCSNR Database nomenclature). The optical, radio and X-ray observations and data reduction are described in Section 2. We discuss the interpretation of the observational data in Section 3 before offering our conclusions in Section 4.

\section{Observations and data reduction}
\subsection{Optical}
The MCELS observations \citep{Smith2006} were taken with the 0.6 m University of Michigan/Cerro Tololo Inter-American Observatory (CTIO) Curtis Schmidt Telescope equipped with a SITE 2048 $\times$ 2048 CCD, producing individual images of $1.35^{\circ} \times 1.35^{\circ}$ at a scale of 2.3$\arcsec$ pixel$^{-1}$. The survey mapped both the LMC ($8^{\circ} \times 8^{\circ}$) and the SMC ($3.5^{\circ} \times 4.5^{\circ}$) in narrow bands covering [\ion{O}{iii}]$\lambda$5007 \AA, H$\alpha$, and [\ion{S}{ii}]$\lambda$6716, 6731 \AA, in addition to matched green and red continuum bands. The survey data were flux calibrated and combined to produce mosaicked images. We extracted cutouts centred on MCSNR~J0527$-$7104 from the MCELS mosaics. Our SNR falls across two regions of differing exposure times in each of the MCELS mosaic's exposure maps. The exposure times of these regions were 2.4~ks and 2.7~ks in the green continuum, red continuum, and H$\alpha$ images, and 4.8~ks and 5.4~ks in the [\ion{O}{iii}] and [\ion{S}{ii}] images. We subtracted the continuum images from the corresponding emission line images, thereby removing the stellar continuum and revealing the full extent of the faint diffuse emission. The continuum subtracted [\ion{S}{ii}] and H$\alpha$ images were used to produce the [\ion{S}{ii}]/H$\alpha$ ratio map shown in Fig. \ref{mcels}a, the colour bar of which has been adjusted to only show regions of enhanced [\ion{S}{ii}] ([\ion{S}{ii}]/H$\alpha\ > 0.4$). MCSNR~J0527$-$7104 is clearly seen, manifest as two bright arcs propagating into the northeast (NE) and southwest (SW) directions. The [\ion{S}{ii}] enhancement to the east of the SNR is related to the N~206 \ion{H}{ii} complex \citep{Henize1956} and is discussed in a separate paper \citep{Kavanagh2012}. In Fig. \ref{mcels}b we show the continuum subtracted [\ion{O}{iii}] image of MCSNR~J0527$-$7104 with [\ion{S}{ii}]/H$\alpha$ contours overlaid, emphasising regions where the ratio is clearly enhanced \citep[\braces{\ion{S}{ii}}/H$\alpha$~$>$~0.67,][]{Fesen1985}. However, due to the green continuum image being heavily contaminated by the foreground star HD 36877 (V = 9.36), the continuum subtraction in this region was somewhat unreliable. To account for this we completely excised the source from the [\ion{O}{iii}] and green continuum images before performing the continuum subtraction. Very faint [\ion{O}{iii}] emission was seen at the leading edges of the enhanced [\ion{S}{ii}] emission regions, notably in the southwestern arc. This provides further evidence for the SNR classification as [\ion{O}{iii}] emission is expected from the cooling zone closest to the SNR shock front.

\subsection{Radio-continuum}
\begin{figure}
\begin{center}
\resizebox{\hsize}{!}{\includegraphics[trim= 3cm 0cm 2.8cm 0cm, clip=true]{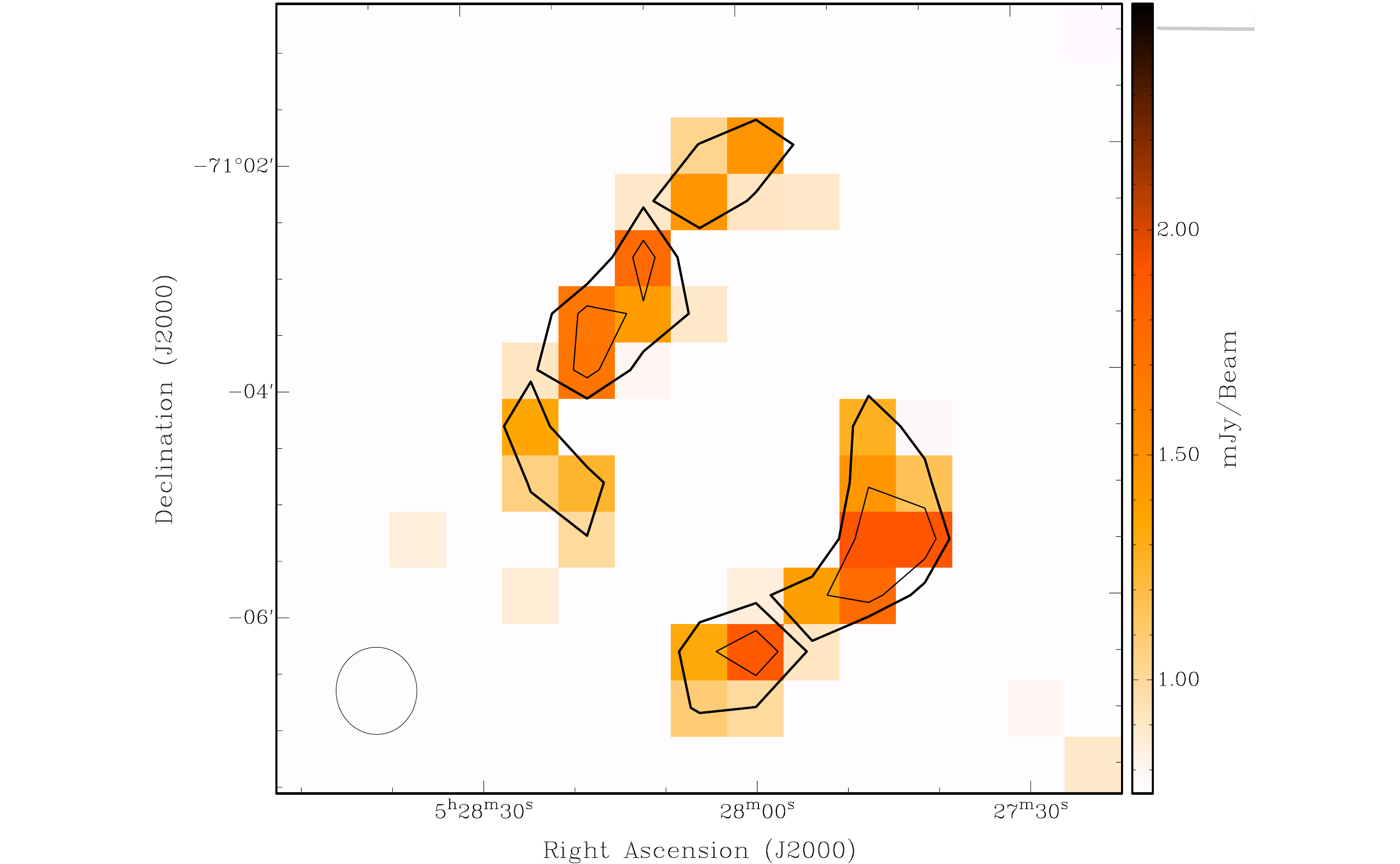}}
\caption{36 cm radio map of MCSNR~J0527$-$7104. Contours are at the 3$\sigma$ and 4$\sigma$ levels ($\sigma$ = 0.25 mJy/beam). The circle in the lower left corner indicates the beamwidth of 43$\arcsec$.}
\label{new-radio}
\end{center}
\end{figure}

\begin{figure}
\begin{center}
\resizebox{\hsize}{!}{\includegraphics[trim= 0.2cm 0.5cm 3.5cm 0.5cm, clip=true]{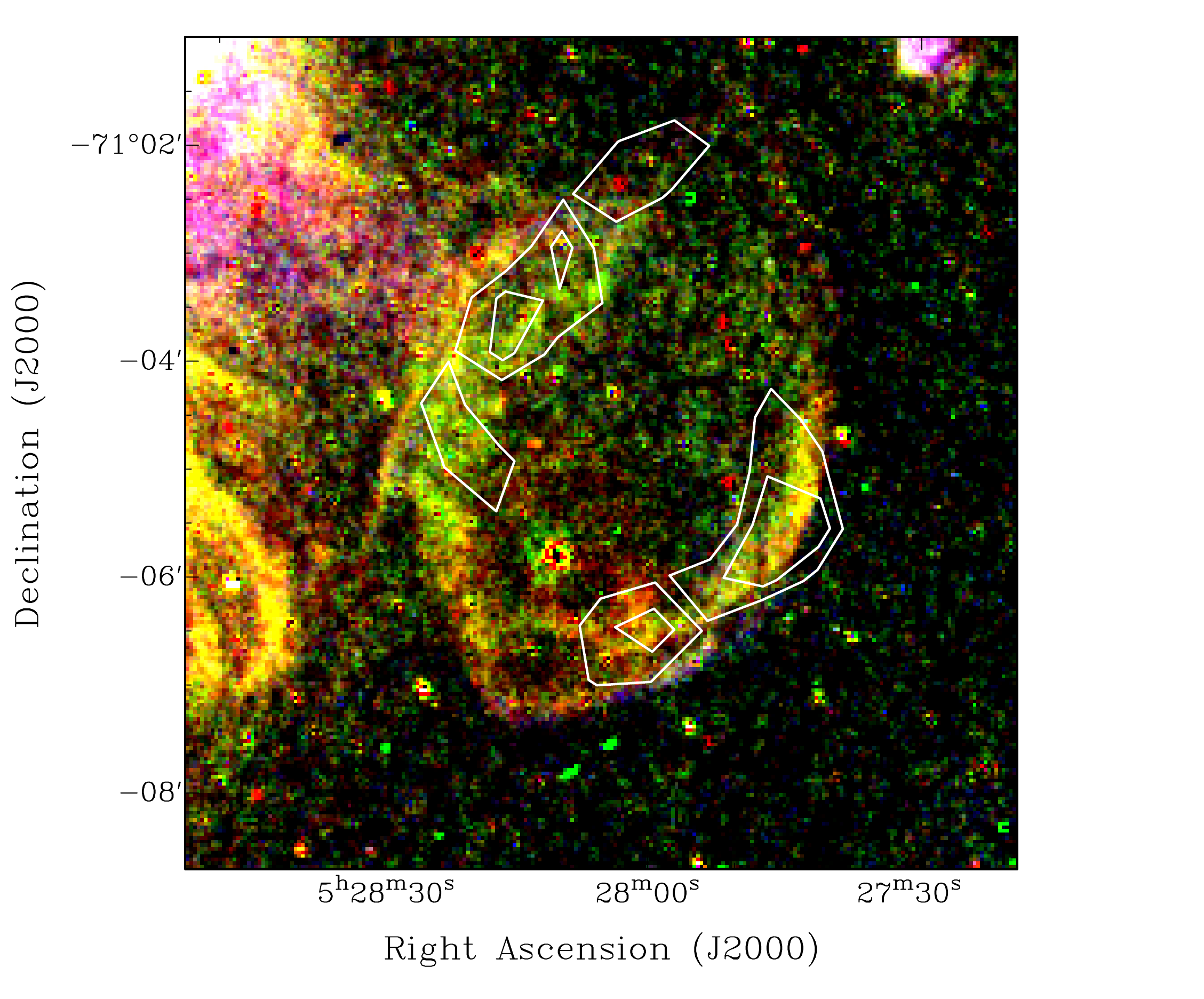}}
\caption{MCELS composite optical image (RGB = H$\alpha$, [\ion{S}{ii}], [\ion{O}{iii}]) of MCSNR~J0527$-$7104 overlaid with 36 cm radio contours determined as in Fig. \ref{new-radio}.}
\label{new-snr}
\end{center}
\end{figure}

We used a mosaic image taken from 36 cm MOST observations \citep[as described in][]{Mills1984} to investigate a radio association with this SNR (Fig. \ref{new-radio}). There were two distinct regions evident in the radio image, located in the northeastern and southwestern areas. This radio morphology is consistent with the long known `bilateral' class of SNRs, characterised by a clear axis of symmetry with low levels of emission along the axis and bright limb emission \cite[see][for examples and descriptions of this morphological class]{Gardner1965,Kesteven1987,Gaensler1998}. At all other radio frequencies, this SNR was well below the detection limit making it one of the weakest SNRs ever detected in the radio. For this reason, polarisation, spectral index and magnetic field studies of MCSNR~J0527$-$7104 were beyond our grasp. In Fig. \ref{new-snr} we plot the radio contours shown in Fig. \ref{new-radio} over the MCELS RGB image, which clearly demonstrates the association of the radio emission with the optical shell of MCSNR~J0527$-$7104.

\subsection{X-rays}
\label{x-rays}
\begin{figure}
\begin{center}
\resizebox{\hsize}{!}{\includegraphics{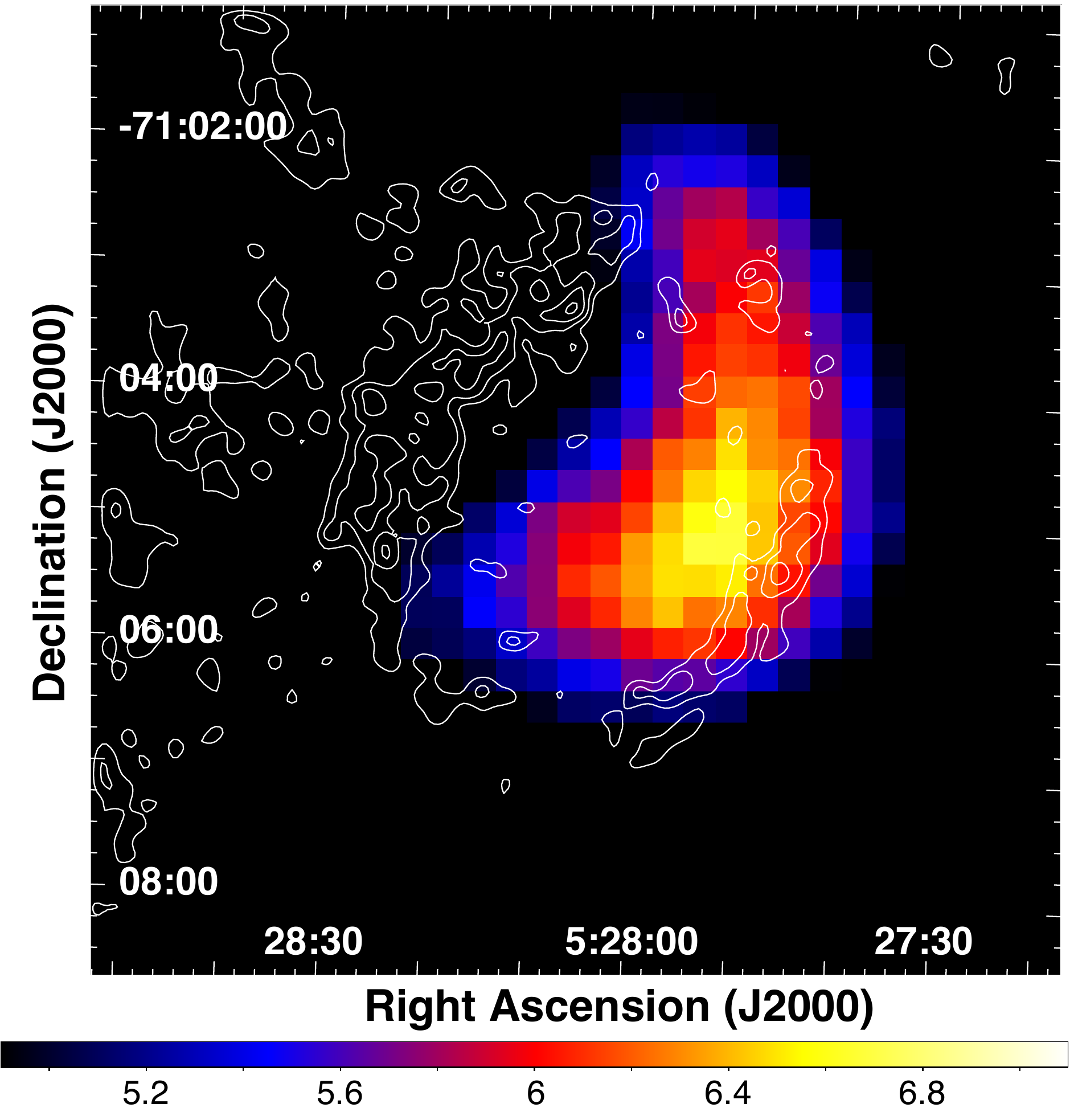}}
\caption{{\it ROSAT} PSPC broad band ($0.5-2$ keV) count rate map of MCSNR~J0527$-$7104. The image has a pixel size of 15$\arcsec$ and is smoothed with a 2$\arcmin$ kernel. The colour bar is in units of $10^{-5}$~counts~pixel$^{-1}$~s$^{-1}$. [\ion{S}{ii}]/H$\alpha$ contours are overlaid, ranging from 0.67 to the maximal value with intermediate
levels of 25, 50, and 75 \% of the maximum.}
\label{rosat}
\end{center}
\end{figure}

The HP99 catalogue was derived from individual PSPC observations, not taking advantage of overlapping
exposures. To examine the relation between X-rays, radio, and optical data, we searched all
observations in the {\it ROSAT} archive covering MCSNR~J0527$-$7104. Unfortunately, we found only two such
observations with identifiers ROR rp300335 and ROR rp300172. The {\it ROSAT} observation with ROR rp300335 was used for the HP99 entry. Both observations were short at 11.3~ks and 13.2~ks, respectively. However, observation rp300172 was split in three parts and the source is seen at an even larger off-axis angle ($\sim50\arcmin$) than in observation rp300335 ($\sim44\arcmin$). Given the low exposure times of each part of  observation rp300172 ($\sim$3 ks, vignetting corrected), the large off-axis angle and corresponding PSF ($\sim3\arcmin$, as opposed to $\sim2\arcmin$ for observation rp300335), we decided to exclude data from observation rp300335 from our analysis as their inclusion would serve to decrease the quality of our results. We produced an exposure corrected image with a pixel size of 15$\arcsec$ from observation rp30033. Due to the limited number of counts ($\sim55$) from the source we did not split the data into different energy bands. Instead, we produced a broad band ($0.5-2$ keV) count rate map of MCSNR~J0527$-$7104, shown in Fig. \ref{rosat}. Taking into account the degraded angular resolution of \textit{ROSAT} at large off-axis angle, we smoothed the image with a Gaussian kernel of 2$\arcmin$. For an easier comparison of X-rays and optical data, we overlaid \ion{S}{ii}/H$\alpha$ line ratio contours derived from Fig. \ref{mcels}a. HP99 determined the extent of this SNR to be $\sim39\arcsec$ but with a very low likelihood for the extent of 0.2. Due to the low number of counts for the SNR in our analysis of observation rp300335 we were prohibited from performing a comparison of the source morphology to the \textit{ROSAT} PSF at the large off-axis angle to confirm the extended nature of the source.

\section{Discussion}
Shell-like optical line emission follows an ellipsoidal morphology, elongated toward a southeast-northwest (SE-NW) axis. Two arcs of enhanced [\ion{S}{ii}]/H$\alpha$ were seen parallel to the major axis, which correlate with faint radio emission at 36 cm. We identified X-ray emission from the west of the SNR in the {\it ROSAT} PSPC data. However, we stress that the remnant was only observed at a large off-axis angle during a shallow observation with {\it ROSAT} (see Section \ref{x-rays}) and, thus, had very limited resolution and sensitivity. Because of this we were prevented from performing a more detailed analysis and the apparent extent of the X-ray emission should be treated cautiously. \\

\par The quality of the radio and X-ray data for this SNR prohibited size determinations in these wavebands. Hence, we determined the dimensions of this SNR from the MCELS data, specifically from the [\ion{S}{ii}] enhanced regions. Given that there is a significant background in the east owing to the N~206 \ion{H}{ii} region, we considered only the regions with clearly enhanced [\ion{S}{ii}] ([\ion{S}{ii}]/H$\alpha\ >$ 0.67), namely the arcs indicated by the contours in Fig. \ref{mcels}b. The size along the NE-SW axis was measured as the distance between the leading edges of the arcs. We determined a size of 4.0$\arcmin$, or $\sim59$ pc at the LMC distance, for this axis. The size of the orthogonal SE-NW axis was somewhat less straightforward given the absence of an optical shell at its extremities. Since the leading edges of the arcs of enhanced [\ion{S}{ii}]/H$\alpha$ were approximately ellipsoidal in shape, we fitted an ellipse to these edges with the minor axis size set to 4.0$\arcmin$, as determined above. We measured a major axis size of $\sim5.1\arcmin$, or $\sim75$ pc at the LMC distance, using this method. However, we emphasise that this estimate is subject to much uncertainty given that we cannot definitively state that this SNR is fully ellipsoidal in shape. We already see some evidence to the contrary if we consider the radio image. In some regions, notably in the north, the arcs of radio emission extend slightly beyond the ellipse representing MCSNR~J0527$-$7104. Thus, our estimate of $\sim5.1\arcmin$ ($\sim75$ pc) in the SE-NW direction is likely lower than the actual SNR dimension. However, without further information a conclusive measurement remains beyond our grasp. Observations by {\it Chandra} or {\it XMM-Newton} would allow the true dimensions of the SNR to be determined, given that the X-ray emitting regions could be traced to much smaller spatial scales. The determined dimensions of MCSNR~J0527$-$7104 of $\sim5.1\arcmin \times\ 4.0\arcmin$ ($\sim75$ pc $\times\ 59$ pc), make it comparable in size to large SNRs such as \object{XMMU~J0541.8$-$6659} \citep{Grondin2012}. However, the MCSNR~J0527$-$7104 is still smaller than the largest confirmed SNRs in the LMC such as \object{SNR~0450$-$70.9} \citep{Matt1985,Williams2004,Cajko2009}, \object{LMC~SNR~J0550$-$6823} \citep{Davies1976,Filipovic1998,Bozzetto2012} and \object{SNR0506-6542} \citep{Klimek2010}. \\

\par If we assume that MCSNR~J0527$-$7104 is in the Sedov phase of its evolution, we can estimate an age from the relation $r(t) = 1.17(E_{0}/\rho_{0})^{1/5}t^{2/5}$ \citep{Woltjer1972}, where $r(t)$ is the radius of the SNR at time $t$, $E_{0}$ is the initial explosion energy, and $\rho_{0}$ is the mass density of the ambient ISM.  \citet{Williams2005} estimated the ISM mass density surrounding the other known SNR in the N~206 \ion{H}{ii} region, \object{SNR~B0532-71.0}, to be $\sim1.2 \times 10^{-25}$~g~cm$^{-3}$, derived from the determined density of the hot gas behind the SNR shock. SNR~B0532-71.0 is also located on the outskirts of N~206, but some $\sim19\arcmin$ ($\sim214$~pc) from MCSNR~J0527$-$7104. \citet{Gorjian2004} reported on {\it Spitzer} observations of the N~206 region, highlighting the dust distribution using 24 $\mu$m, 70 $\mu$m, and 160 $\mu$m images. These images suggested that the ISM around SNR B0532-71.0 is slightly denser than the ISM around MCSNR~J0527$-$7104. It is not ideal to adopt the ambient ISM mass density of SNR B0532-71.0 for that of MCSNR~J0527$-$7104, especially given the calculations are quite sensitive to this parameter. However, it allows us to at least suggest an age for our SNR. For a canonical initial input energy of $10^{51}$~erg, we determined an age of $\sim$25 kyr for MCSNR~J0527$-$7104 using the Sedov relation. We again stress that this age determination is very sensitive to the adopted ambient density, as well as the initial explosion energy. Using these parameters and our age estimate, we can rearrange the Sedov relation (setting the units as the input parameter values of $\rho_{0} = 1.2\times10^{-25}$ g cm$^{-3}$ and $E_{0}=10^{51}$) to highlight the dependence of the SNR age on the ambient mass density and initial explosion energy as

\begin{equation}
t \approx 25 \left(\dfrac{E_{0}}{10^{51} \rm{erg}}\right)^{-\frac{1}{2}} \left(\dfrac{\rho_{0}}{1.2\times10^{-25} \rm{g cm^{-3}}}\right)^{\frac{1}{2}} \rm{kyr}.
\end{equation}

So, for example, an increase in the ambient mass density to double our adopted value results in an increase of $\sim$10 kyr in age to $\sim35$ kyr. A similar increase in age is seen if the initial explosion energy is halved. A combination of both of these examples leads to a doubling of the age estimate to $\sim50$ kyr. Hence, our determined age should be treated with the utmost caution. \\

\par Using the adopted ambient density and explosion energy we can also estimate the time at which this SNR will enter the radiative phase ($t_{\rm{rad}}$) using the relation of \citet{Vink2012},

\begin{equation}
t_{\rm{rad}} = 1.5\times10^{-13}\left(\dfrac{\xi E_{0}}{\rho_{0}}\right)^{\frac{1}{3}},
\end{equation}

where $\xi$ is a dimensionless constant depending on the adiabatic index, which is 2.026 for a monatomic gas with $\gamma = 5/3$. We determine that MCSNR~J0527$-$7104 will enter the radiative phase at $\sim$122~kyr. While this estimate seems to verify our original assumption that MCSNR~J0527$-$7104 is indeed in the Sedov phase of its evolution, we again stress the dependence on our assumed ambient mass density and initial explosion energy values and the resulting uncertainty in age and $t_{\rm{rad}}$. To obtain a more robust age estimate would require more observational data than is currently available. High quality X-ray data from {\it XMM-Newton} and/or {\it Chandra} would certainly be beneficial. By determining the physical characteristics of the hot gas in the SNR from X-ray spectral analysis, the age of the SNR could be inferred \citep[see][for example]{Maggi2012}. Alternatively, optical echelle data could provide a measurement of the expansion velocity of the SNR shell, from which an age could be determined \citep[see][for example]{Williams2004}. \\

\begin{figure*}
\begin{center}
\resizebox{\hsize}{!}{\includegraphics{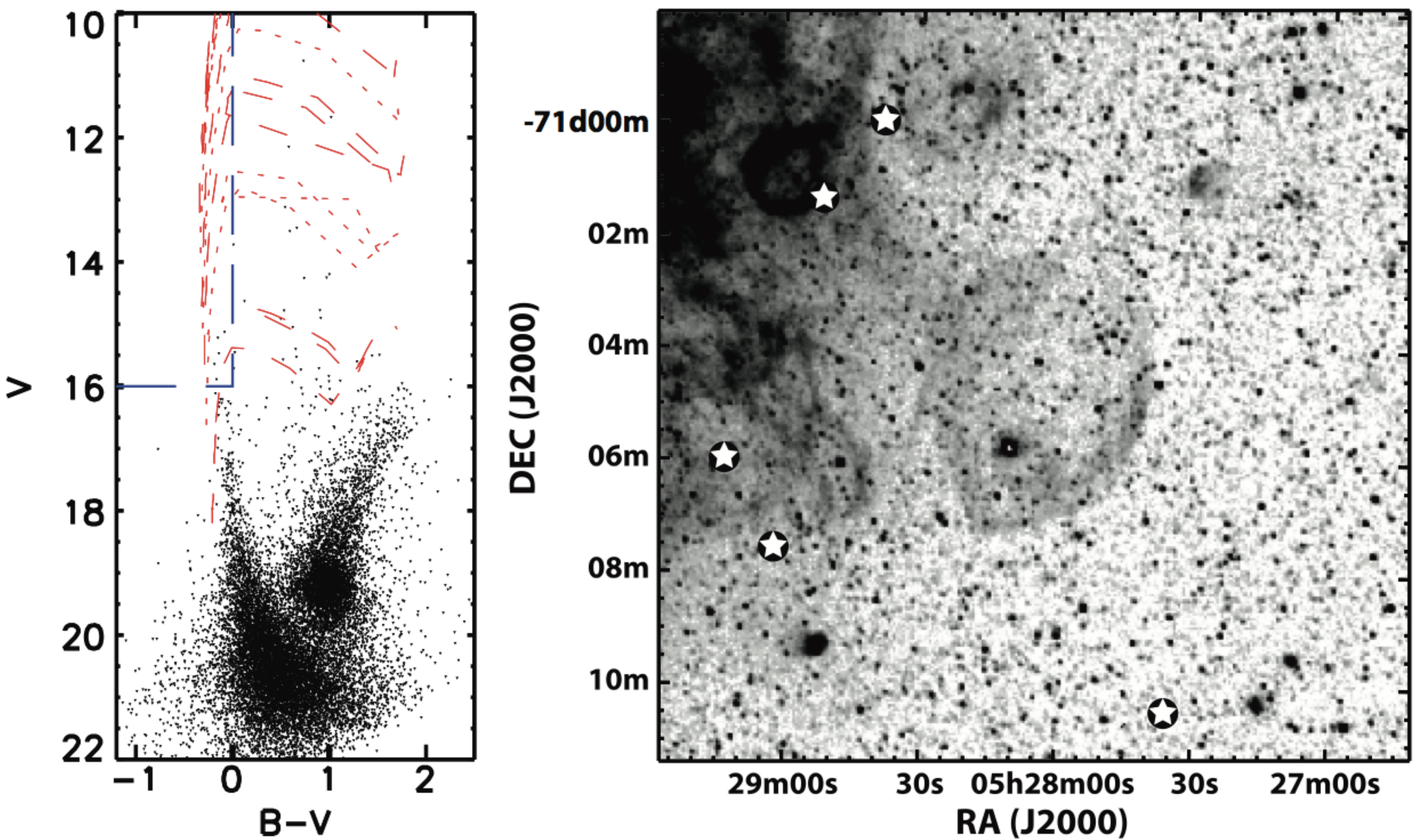}}
\caption{Photometric identification and location of candidate blue high-mass stars within 100 pc of MCSNR~J0527$-$7104. Left panel: $B-V$, $V$ colour-magnitude diagram of stars within 100 pc of MCSNR~J0527$-$7104, taken from the Magellanic Cloud Photometric Survey \citep{Zaritsky2004}. Also shown are stellar evolutionary tracks from \citet{Lejeune2001} with, from bottom to top, the long dashed red lines showing the 5 M$_{\sun}$, 15 M$_{\sun}$, 25 M$_{\sun}$, and 60 M$_{\sun}$ tracks, and the short dashed red lines showing the 10 M$_{\sun}$, 20 M$_{\sun}$, and 40 M$_{\sun}$ tracks. The selection criteria for identification of blue high-mass stars is indicated by the dashed blue lines. Right panel: Location of the candidate blue high-mass stars in relation to MCSNR~J0527$-$7104 marked on the MCELS H$\alpha$ image.}
\label{phot}
\end{center}
\end{figure*}

\par
\begin{figure}
\begin{center}
\resizebox{\hsize}{!}{\includegraphics[trim= 2cm 1.5cm 4cm 14cm, clip=true]{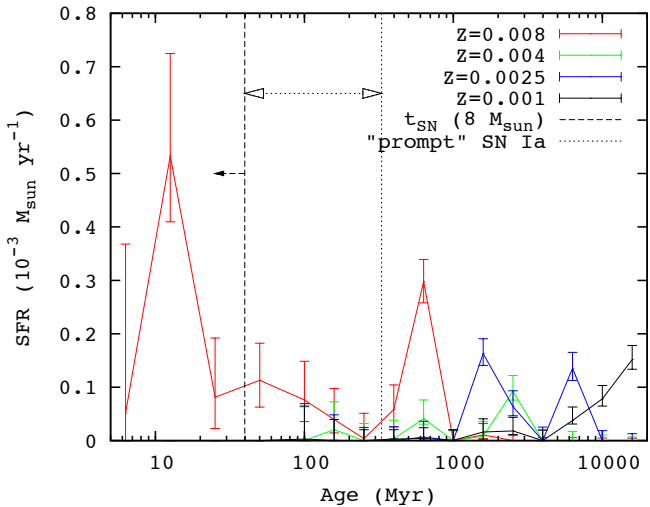}}
\caption{Average SFH of the $12\arcmin \times 12\arcmin$ cells, centred on MCSNR~J0527$-$7104, from the data of \citet{Harris2009}. The SFH is broken into four metallicity bins as indicated in the plot key. Also shown are the time ranges from which we expect either core collapse (indicated by the dashed arrow) or `prompt' type Ia (indicated by the dotted double sided arrow) progenitors to be formed. The peak at 12.5 Myr reflects the higher recent star formation activity seen only in the northeastern cell which contains N~206. Another episode of star formation is seen $\sim630$ Myr ago. }
\label{sfh}
\end{center}
\end{figure}

 Given the propensity of high-mass stars to form in clusters, we can assess the likely supernova (SN) mechanism that produced MCSNR~J0527$-$7104 based on the surrounding high-mass stellar population, or lack thereof. We used data from the Magellanic Cloud Photometric Survey \citep[MCPS,][]{Zaritsky2004} to produce a colour-magnitude diagram (CMD) for the survey entries within 100 pc ($6.6\arcmin$) of MCSNR~J0527$-$7104 (Fig. \ref{phot} left panel). Candidate blue stars more massive than $\sim 8$ M$_{\sun}$ were identified using the appropriate photometric selection criteria ($V < 16, B-V < 0$), yielding 5 candidate main sequence B-stars. Interestingly, four of the five candidates were located to the east of MCSNR~J0527$-$7104 toward the N~206 \ion{H}{ii} region. There was, however, no candidate in the immediate vicinity of the SNR. N~206 is known to harbour several high-mass stellar associations \citep{Massey1995,Bica1999} with ages $< 10$~Myr \citep{Gorjian2004,Kavanagh2012}, the closest of which is $\sim150$~pc from our SNR. In addition, the region is known to be actively forming stars \citep{Romita2010}, a picture that is echoed by the analysis of \citet{Harris2009}, who reconstructed the spatially resolved star formation history (SFH) of the LMC. The star formation rate of N206 has been relatively high for $\sim 50$~Myr, with a notable surge in the last $\sim12.5$~Myr. MCSNR~J0527$-$7104 was at the centre of four $12\arcmin \times 12\arcmin$ cells in the SFH map of \citet{Harris2009}. The average SFH from these cells is shown in Fig. \ref{sfh}. The peak at 12.5~Myr reflects the higher recent star formation activity seen only in the northeastern cell, in N~206. Another episode of star formation was seen $\sim630$~Myr ago, about the timescale for a `delayed' SN Ia to go off. However, as stressed by \citet{Badenes2009}, care should be taken in regions with inhomogeneous stellar populations on short scale. If, due to dynamical ejection or SN kick processes, a runaway star with a modest ejection/kick velocity of $\sim 30$ km s$^{-1}$ \citep[][and references therein]{Perets2012} was ejected from the N~206 stellar associations, it could travel $\sim150$~pc in $<5$~Myr. In an SN kick scenario, this travel time in addition to the time until a binary companion $>20$~M$_{\sun}$ undergoes an SN event is well within the lifetime of the lowest mass core-collapse SN progenitors. Thus, even though the old local stellar population favours a type Ia event, the possibility of a core-collapse SN by a runaway massive star remains plausible. Only high quality X-ray spectra with sufficient counts to determine plasma abundances will resolve this issue provided X-ray emission from SN ejecta can be detected and separated from swept-up ISM emission.

\section{Conclusions}
We have detected significant H$\alpha$, [\ion{S}{ii}], and [\ion{O}{iii}] optical line emission, characteristic of an optical SNR, in the region of the X-ray identified candidate SNR [HP99] 1234. In addition, very faint radio emission at 36 cm is detected coincident to the optical line emission which is consistent with a `bilateral' SNR morphology. At all other radio frequencies this SNR was well below the detection limit making it one of the weakest SNRs ever detected in the radio. The enhanced [\ion{S}{ii}] and faint radio emission were found to be shell-like with an ellipsoidal morphology. In contrast, the X-ray emission was concentrated towards the west of the SNR. However, due to its location at a large off-axis angle in a shallow \textit{ROSAT} observation, the extent of the X-ray emission should be treated cautiously. The dimensions of the SNR, determined from the higher quality optical data, were found to be $\sim5.1\arcmin\ \times 4.0\arcmin$ ($\sim75$ pc $\times\ 59$ pc). However, the size estimate along the major axis was somewhat uncertain due to an absence of optical and radio emission at its extremities and the sensitivity and resolution of the X-ray data being too poor to perform a reliable measurement. We assumed the SNR to be in the Sedov phase of its evolution and determined an age estimate of $\sim$25 kyr for a canonical initial explosion energy of $10^{51}$ erg and an ambient mass density of $1.2 \times 10^{-25}$ g cm$^{-3}$, measured for another SNR in the N~206 region. Additionally, we determined that for the adopted parameters, MCSNR~J0527$-$7104 will enter the radiative phase of its evolution after $\sim122$ kyr, validating our initial assumption that the SNR is in the Sedov phase. However, we stress the sensitivity of the age and radiative time estimates to the assumed explosion energy and ambient mass density. We assessed the SN explosion mechanism responsible for our SNR based on the nearby stellar population, showing that both core-collapse and Type Ia scenarios are possible. \\

\par With the detection of radio emission, X-ray emission and significant H$\alpha$, [\ion{S}{ii}], and [\ion{O}{iii}] line emission with a [\ion{S}{ii}]/H$\alpha$ flux ratio $>0.4$, this SNR satisfied all the criteria for SNR classification and we assign it the identifier of MCSNR~J0527$-$7104. However, issues such as the uncertain age and size of the SNR, the true explosion mechanism for the SNR and the explanation for the very faint radio emission persist. High quality X-ray images and spectra obtained with {\it XMM-Newton} and/or {\it Chandra} would provide answers to these outstanding questions.

\begin{acknowledgements}
We wish to thank the anonymous referee for their constructive suggestions to improve the paper. P.K. and P.M. are funded through the BMWI/DLR grants FKZ 50 OR 1009 and 50 OR 1201, respectively. M. S. acknowledges support by the Deutsche Forschungsgemeinschaft through the Emmy Noether Research Grant SA 2131/1. The MCELS data are kindly provided by R.C. Smith, P.F. Winkler, and S.D. Points. The MCELS project has been supported in part by NSF grants AST-9540747 and AST-0307613, and through the generous support of the Dean B. McLaughlin Fund at the University of Michigan. The National Optical Astronomy Observatory is operated by the Association of Universities for Research in Astronomy Inc. (AURA), under a cooperative agreement with the National Science Foundation. This research has made use of the SIMBAD database operated at CDS, Strasbourg, France.

\end{acknowledgements}

\bibliographystyle{aa}
\bibliography{refs.bib}

\end{document}